\RequirePackage{ifpdf}
\documentclass{PoS}
\pdfoutput=1
\usepackage{caption}
\usepackage{subcaption}
\usepackage{amsmath}

\newcommand{\be}{\begin{equation}}
\newcommand{\ee}{\end{equation}}
\newcommand{\bea}{\begin{eqnarray}}
\newcommand{\eea}{\end{eqnarray}}
\newcommand{\non}{\nonumber}
\newcommand{\bie}{\begin{small} \begin{itemize}}
\newcommand{\ie}{\item}
\newcommand{\eie}{\end{itemize} \end{small}}
\newcommand{\bref}{\begin{flushright} \begin{small} \color{red} }
\newcommand{\eref}{\end{small} \end{flushright}}

\def\s#1{{\scriptscriptstyle #1}}

\title{Gauge fixing and the gluon propagator in renormalizable xi gauges}

\ShortTitle{Gauge fixing and gluon propagator in $R_\xi$ gauges}

\author{\speaker{Pedro Bicudo}
\\
        CFTP, Instituto Superior T\'{e}cnico, Universidade de Lisboa, Av. Rovisco Pais, 1049-001 Lisboa, Portugal\\
        E-mail: \email{bicudo@tecnico.ulisboa.pt}}

\author{Daniele Binosi \\
European Center for Theoretical Studies in Nuclear Physics and Related Areas (ECT*) and Fondazione Bruno Kessler, Italy\\
        E-mail: \email{binosi@ectsar.eu}}

\author{Nuno Cardoso \\
NCSA, University of Illinois, Urbana IL 61801, USA\\
        E-mail: \email{ncardoso@illinois.edu}}

\author{Orlando Oliveira \\
CFisUC, Department of Physics, University of Coimbra, P-3004 516 Coimbra, Portugal\\
        E-mail: \email{orlando@teor.fis.uc.pt}}

\author{Paulo J. Silva \\
CFisUC, Department of Physics, University of Coimbra, P-3004 516 Coimbra, Portugal\\
        E-mail: \email{psilva@teor.fis.uc.pt}}

\abstract{Covariant $R_\xi$ gauge fixing is notoriously difficult for large lattice volumes, large $\xi$ and small $N_c$. We thoroughly test different convergence techniques, which allows the gauge fixing of lattice configurations with a total volume of (3.25 fm)$^4$, up to $\xi=0.5$. We are able  to study the gluon propagator in the infrared region and its dependence on the gauge fixing parameter $\xi$. As expected, the longitudinal gluon dressing functions stay constant at their tree-level value $\xi$. Similar to the Landau gauge, the transverse $R_\xi$ gauge gluon propagators saturate at a non-vanishing value in the deep infrared for all values of $\xi$ studied. We compare with very recent continuum studies and perform a simple analysis of the found saturation with a dynamically generated effective gluon mass.}

\FullConference{The 33rd International Symposium on Lattice Field Theory\\
		14 -18 July 2015\\
		Kobe International Conference Center, Kobe, Japan*}

\begin{document}

\section{Introduction}

\begin{figure}[t]
\begin{center}
\includegraphics[width=0.45\textwidth]{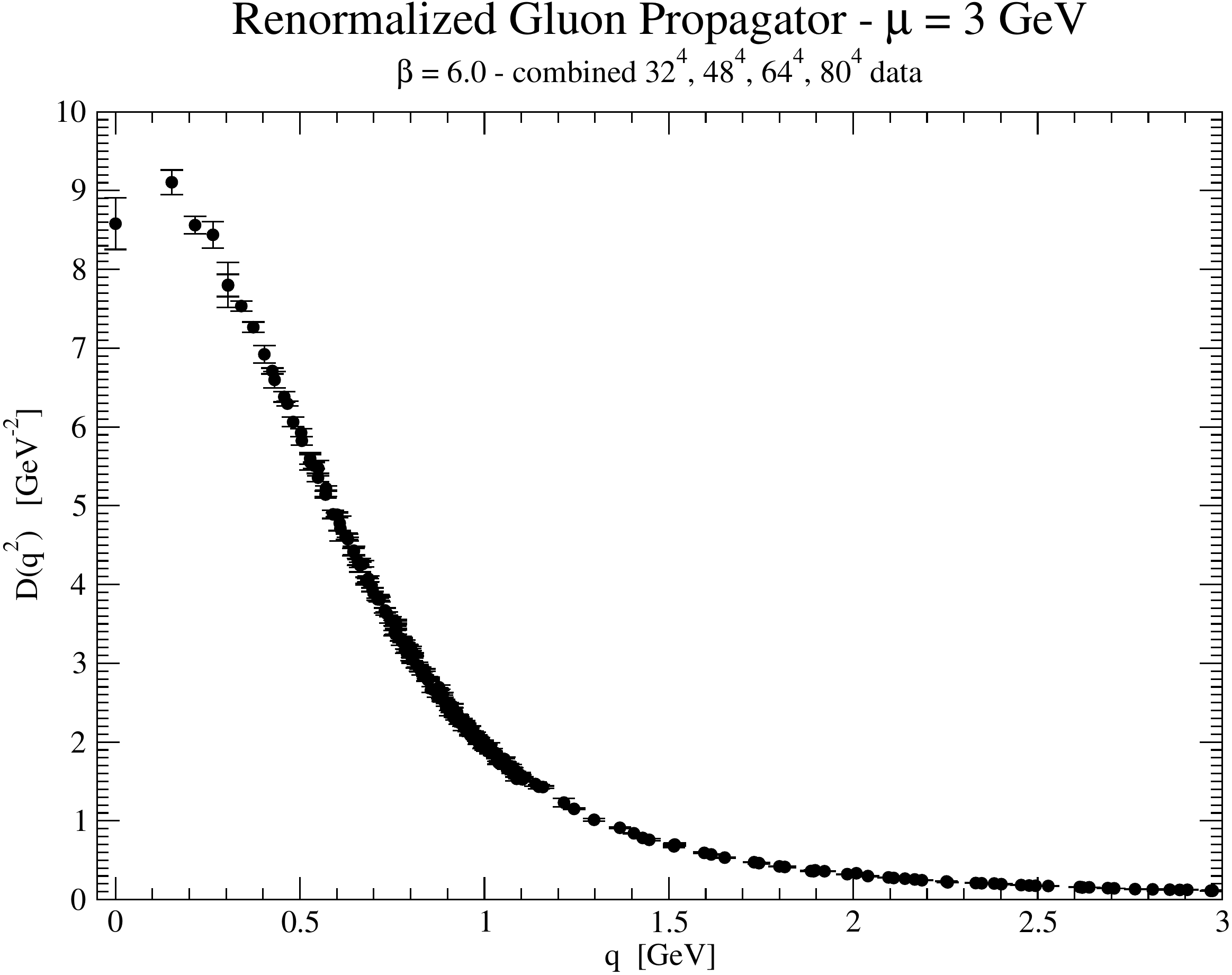}
\qquad
\includegraphics[width=0.45\textwidth,trim={0 10.0cm 0 0},clip]{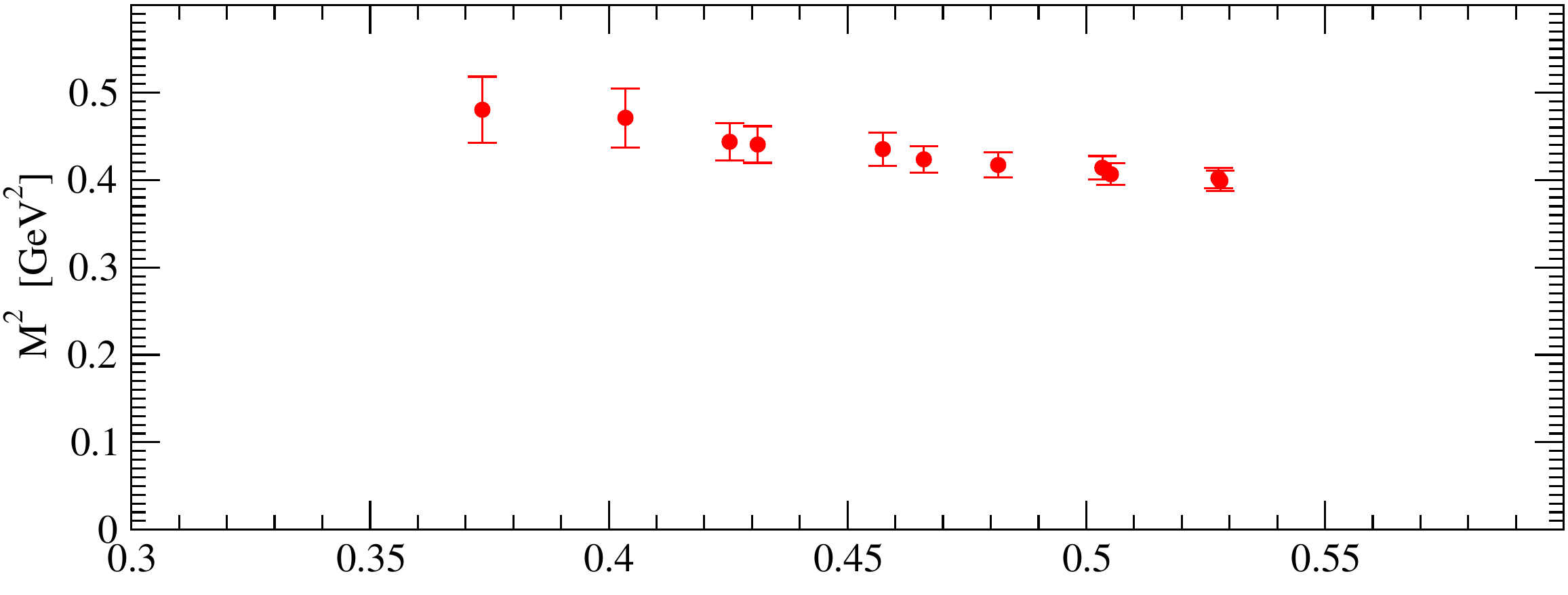}
\end{center}
\caption{ 
\label{fig:massLandau}
(left) Gluon propagator and (right) gluon mass computed in the Landau gauge \cite{Oliveira:2010xc}.}
\end{figure}

Dynamical mass generation in QCD and hadrons accounts for $\sim$ 98\% of the visible mass in the universe.
Most of the mass of the light quarks is due to chiral symmetry breaking. An effective gluon mass (as the photon mass in a superconductor) is a possible signal of confinement \cite{Cornwall:1981zr,Aguilar:2004sw}. 
If it exists, the gluon mass should appear in the gluon propagator.
However Green's functions (propagators and vertices) depend on the gauge fixing.
In the case of the Landau gauge,
the  transverse part of the gluon propagator saturates in the IR, as illustrated in Fig. \ref{fig:massLandau}.
This can be interpreted as an evidence for a dynamically generated mass
\cite{Oliveira:2010xc}. 
While there is yet no unique way to define the gluon mass, in general the gluon mass $m^2 (p^2)$ is a monotonically decreasing function,  power-law suppressed in the UV.
The question is, what happens in other gauges?  So far , we don't really know.

Here we study  in Lattice QCD the {\em well known} renormalizable-$\xi$ covariant gauges (including Landau gauge  $\xi=0$ and Feynman gauge $\xi=1$ ). 
Reliable lattice calculations in renormalizable-$\xi$ ($R_\xi$) covariant gauges have not been systematically pursued previously. 
$R_\xi$ gauge fixing (GF)  lattice implementation has in fact proven to be quite complicated \cite{Giusti:1996kf}.
Recently some success was achieved \cite{Cucchieri:2009kk}, however one still encounters significant convergence problems in realistic lattices, i e  for larger GF parameter $\xi$ and lattice size, and smaller number of colours $N_c$ and lattice coupling $\beta$. 

We report on our success \cite{Bicudo:2015rma} to GF in a realistic lattice and present the SU(3) gluon propagator in $R_\xi$ gauges for a relatively large lattice volume (3.25 fm)$^4$ and a GF parameter up to $\xi=0.5$.


\section{$R_\xi$ gauges framework and gauge fixing algorithm}

\subsection{$R_\xi$ gauges}

In the continuum, 
$R_\xi$ GF is achieved by adding to the SU($N_c$) Yang-Mills action the term,
\be
S_\s{\mathrm{GF}}=\int\!\mathrm{d}^4x\,\left[b^m \Lambda^m-{\xi \over 2}(b^m)^2\right] \ ,
\ee
where $\xi$ is the GF parameter, $b^m$ are Nakanishi-Lautrup multipliers and  $\Lambda^m=\Lambda^m[A]$ is the GF condition. 
Going on-shell,  $\xi b^m=\Lambda^m$, and the GF action takes a Gaussian form,
\be
S_\s{\mathrm{GF}}={ 1 \over 2 \xi} \int\!\mathrm{d}^4x\,( \Lambda^m)^2 \ .
\label{eq:GFaction}
\ee
$R_\xi$ gauges are obtained when the linear condition  is chosen
\be
\Lambda^m=\partial^\mu A^m_\mu \ .
\label{eq:GFlinear}
\ee 
When the gluon propagator is decomposed in transverse / longitudinal components,
\be  
	\Delta_{\mu\nu}(q)=(g_{\mu\nu}-q_\mu q_\nu/q^2)\Delta_\s{\mathrm{T}}(q^2)+(q_\mu q_\nu/q^2)\Delta_\s{\mathrm{L}}(q^2)\ ,
  \label{eq:Rxiprop}
\ee
Slavnov-Taylor identities ensure that $q^2\Delta_\s{\mathrm{L}}=\xi$ to all orders.

In the lattice, we use the Wilson action and the gauge links $U_\mu$, are related to the gauge fields,
\be
  A_\mu (x +  \widehat{e}_\mu / 2 ) = \left. \frac{U_{\mu}(x) - U^\dagger_{\mu}(x)}{2i g_0} \right|_{\mathrm{traceless}} \ .
\label{eq:traceless}
\ee
In $R_\xi$ gauges, besides the usual integration over the link variables $U_\mu(x)$ , one has to integrate over the fields $\Lambda = \sum_m \Lambda^m  t^m$, 
where each $ \Lambda^m$ is a {\em Gaussian distribution with variance  $ {2 N_c \over \beta} \, \xi$ }.

\begin{figure}[t]
\begin{subfigure}{.45\linewidth}
\includegraphics[width=1.05\textwidth, height=143pt]{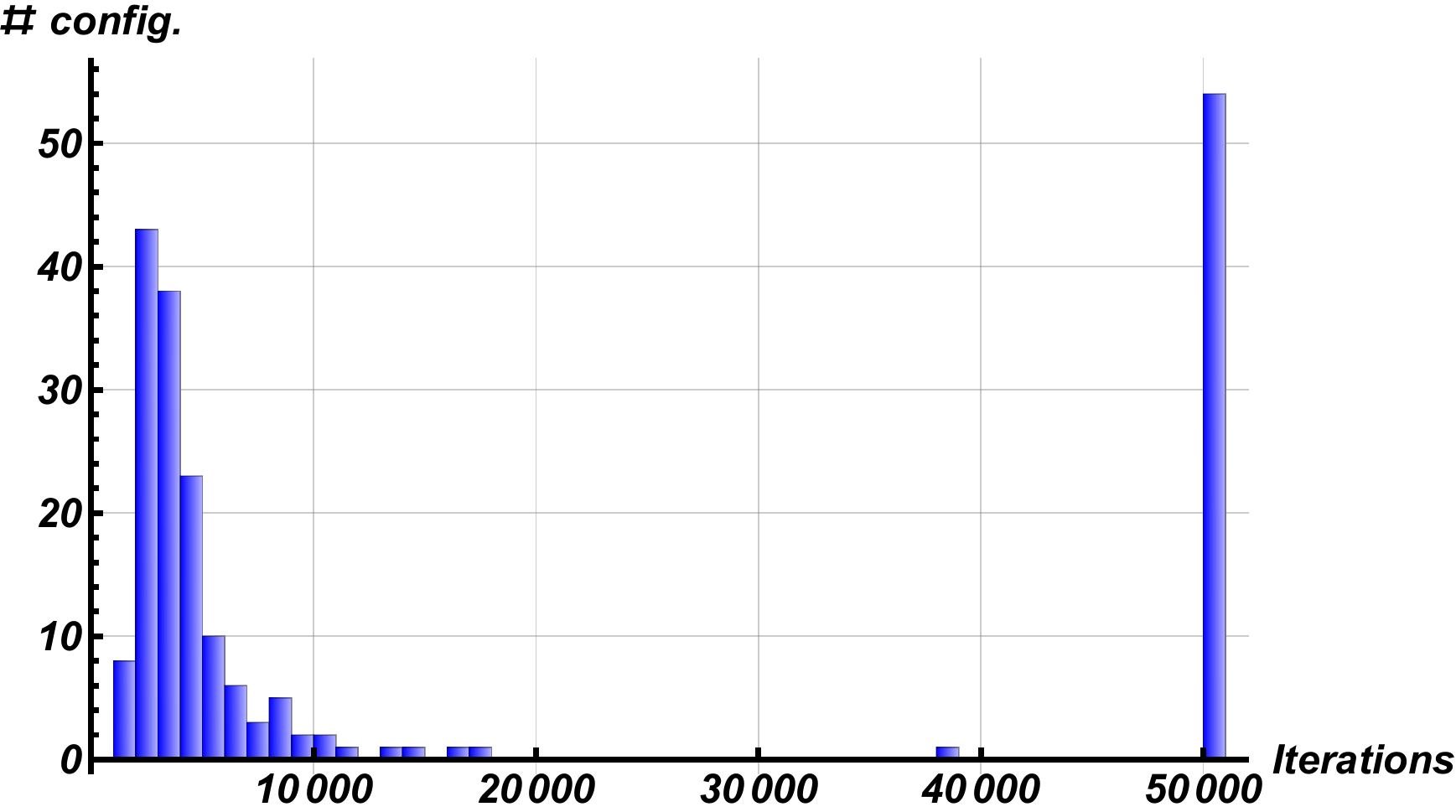}
\caption{
\label{fig:FFT}
Histogram of the number of iterations necessary to converge the $R_\xi$ GF, with  FFT steepest descent minimization of the functional $\theta$. We illustrate 200 uncorrelated configurations  for $\xi=0.3$ and $L=32$. After 50000 iterations we stop FFT and cycle through other GF techniques, OVR, STR, RGT  until convergence is reached.}
\end{subfigure}
\qquad
\begin{subfigure}{.47\linewidth}
\includegraphics[width=1.05\textwidth]{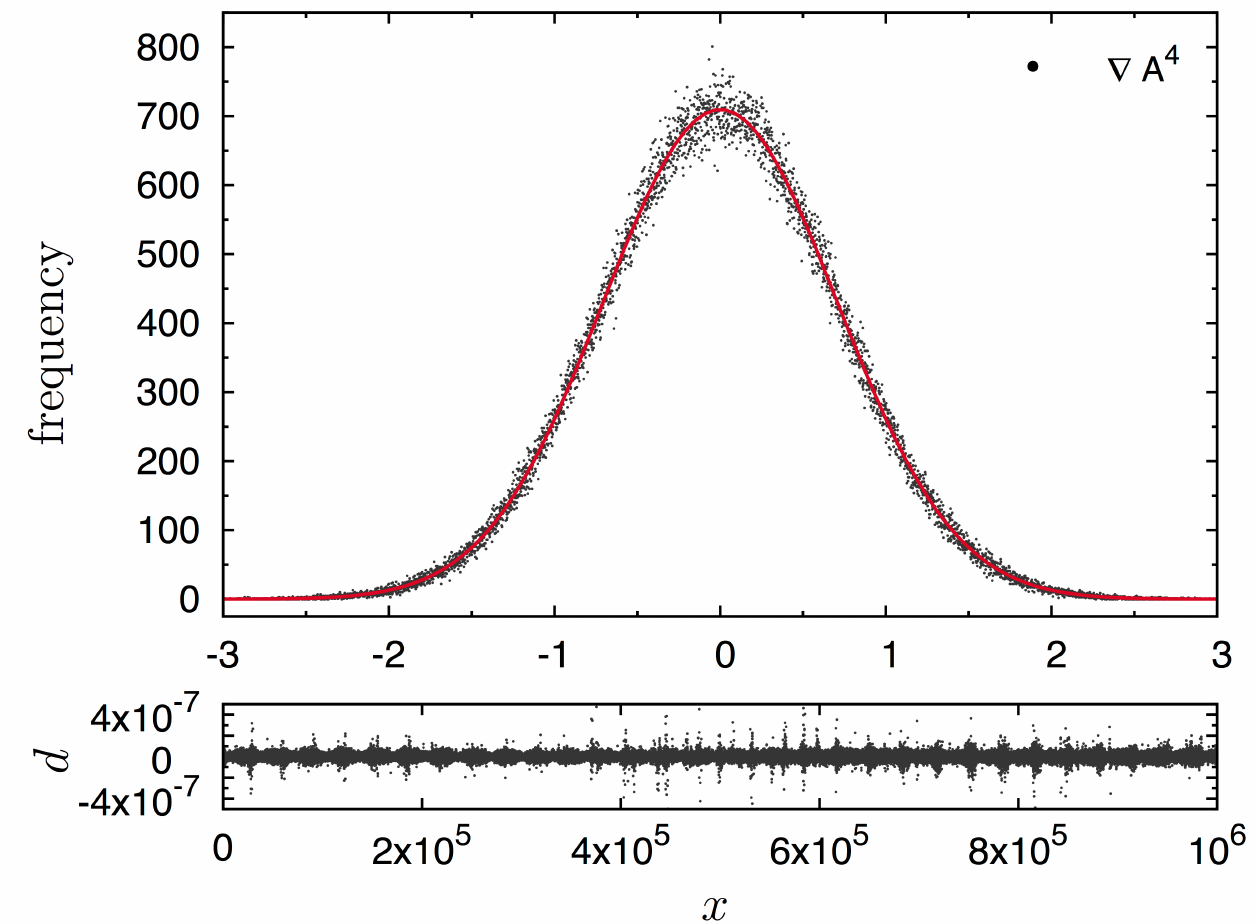}
\caption{
\label{fig:success}
({\em top})
 The $32^4$ values of $\nabla \! \cdot \! A^m$ evaluated for a configuration gauge fixed at $\xi=0.5$, grouped in 5000 bins,  compared with the Gaussian $\Lambda^m$ with standard deviation $\sqrt{\xi} \simeq 0.316$. 
 ({\em Bottom}) Plot of $d=\nabla \! \cdot \! A^4-\Lambda^4$; the two distributions coincide within $\sqrt \theta$ precision.}
\end{subfigure}
\caption{ Our approach to achieve the $R_\xi$ GF within the desired precision of $\theta < 10^{-15}$.}
\end{figure}

The procedure for GF requires to gauge rotate all link variables \cite{Giusti:1996kf,Cucchieri:2009kk},
\be
U_{\mu}(x) \rightarrow g(x)U_{\mu}(x)g^{\dag}(x+\widehat e_{\mu})
\ee
where $g$ are elements of the SU($N_c$)  gauge group. 

In the Landau gauge case, which is the $\xi\to0$ limit of the $R_\xi$ gauges studied here, the GF is implemented minimizing the functional, $ -\mbox{Re} \, \mbox{Tr}\, \sum_{x, \mu}g(x) U_{\mu}(x) g^{\dagger}(x+ \widehat e_\mu)\, ,
 $
since the minimization ofr any functional of $A_\mu(x)$ directly leads to the condition \mbox{$\nabla\!\cdot\! A^m=0$}.

\subsection{$R_\xi$ gauge fixing algorithm}

\begin{figure}[t]
\begin{subfigure}{.45\linewidth}
\includegraphics[scale=0.9,trim={0 0 8.5cm  0},clip]{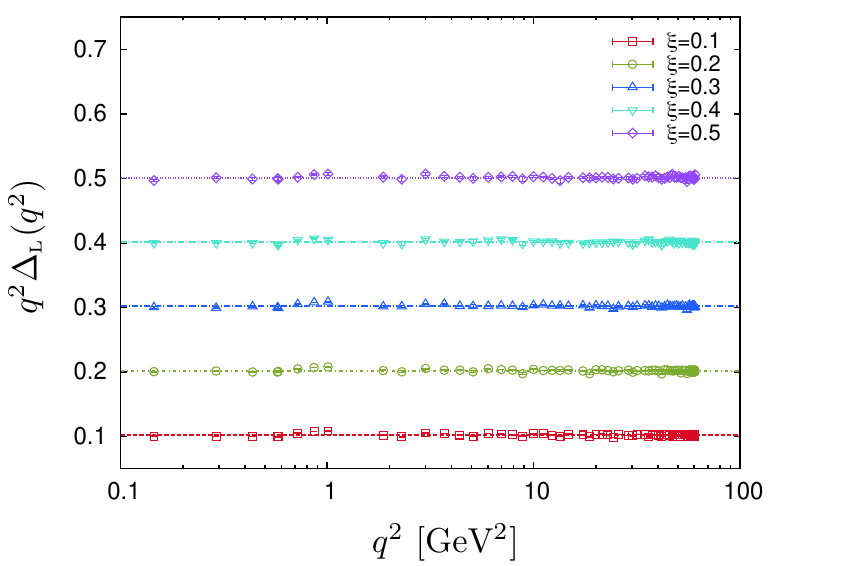}
\caption{ 
\label{fig:longdress}
Gluon longitudinal $R_\xi$  dressing function $q^2\Delta_\s{\mathrm{L}}$, compared to the theoretical value $\xi$.}
\end{subfigure}
\quad
\begin{subfigure}{.45\linewidth}
\includegraphics[scale=0.9,trim={8.5cm 0 0 0},clip]{gl-longit-tr-dress.pdf}
\caption{
\label{fig:transdress}
Gluon transverse $R_\xi$  dressing function $q^2\Delta_\s{\mathrm{T}}$,  (renormalized at $\mu=4.317$ [GeV] ).}
\end{subfigure}
\caption{ Guon propagator dressing functions.}
\end{figure}

In the general case of a non-vanishing~$\xi$,  a more complicated functional than the one of Landau GF needs to be minimized. 
In practice, the gauge transformation, \   $g = \prod_j \delta g_j$, \ is built as a product of a sequence of infinitesimal gauge transformations. 
For each infinitesimal transformation, \  $\delta g_j  = 1+ i \sum_m w^m \,  t^m$,  \ \ we minimize the $R_\xi$ functional with respect to $w^m$.   
We directly utilize the linear gauge condition of Eq. (\ref{eq:GFlinear}), to define our functional $\theta$, depending on $U_\mu(x)$ via Eq. (\ref{eq:traceless}),
\bea
 \theta &=& \frac{1}{ N_c L^4}\sum_x \mathrm{Tr}\, [\Delta(x) \Delta^{\dagger}(x)] \ ,
\non \\
 \Delta(x) &=& \sum_\mu g_0 \Big[ A_\mu (x + \hat{e}_\mu / 2 )- A_\mu (x -  \hat{e}_\mu / 2 ) \Big] - \Lambda(x) \ .
\label{eq:omega}
\eea
When $\theta \to 0$, then  $\Delta(x) \to 0$ and we reach the desired gauge condition ~$\nabla\!\cdot\! A^m=\Lambda^m$ of Eq. (\ref{eq:GFlinear}).

Thus, choosing~$w^m = \alpha \Delta^m$, where $\alpha$ is a relaxation parameter to be optimized, should reduce $\Delta$ with a {\em steepest descent method}.
 Our goal is to converge to a vanishing $\Delta(x)$ in all lattice points $x$. 
Based on our experience of studying gluon propagators with Landau GF, we aim at a very small $\theta < 10^{-15}$.
However, this turns out, in practice, to be very difficult in a realistic lattice!

We opt to extend {\em  our fully parallel and  very fast GPU codes for Landau GF} \cite{Cardoso:2012pv}, for a thorough optimization of all possible techniques to minimize $\theta$.
We combine three different GF techniques optimized in the Landau case \cite{Cardoso:2012pv},
\bie
\ie
the Fast Fourier Transform -  accelerated steepest descent  (FFT),
\ie
Over Relaxation (OVR)
\ie
and Stochastic Relaxation (STR).
\eie
Each one of theses techniques, with an optimized convergence parameter $\alpha$ has a similar $R_\xi$ GF success rate of only $\sim$ 75\% for $\xi=0.3$ \  and $\sim$ 40\% for $\xi=0.5$.  This is illustrated in Fig. \ref{fig:FFT}

After a large number of tests performed in our GPU servers,  we finally find a solution. Cycling through FFT, OVR and STR, for our hardest case of $\xi=0.5$, we increase the convergence success rate up to $\sim$90\% .
Finally, for the remaining 10\% cases,  we perform a random gauge transformation (RGT), and restart the combined algorithm, till convergence is total
\cite{Bicudo:2015rma,preparation}, see Fig. \ref{fig:success}.


\section{Results and comparison with continuum studies}

\subsection{Results for the gluon propagator}

We now compute the gluon propagator for $\beta=6.0$ in a $32^4$ lattice.
Linear covariant gauges on the lattice requires, not only an integration over the ensemble of configurations $U_\mu(x)$ \cite{ Cardoso:2011xu}, but an additional integration over the $N_c^2-1$ Gaussian distributed $\Lambda^m$ fields.
For the $\Lambda$ integration, we consider 50 different  $\Lambda$'s for each configuration $U_\mu$.
Then, for each configuration of $U_\mu(x)$ and $\Lambda(x)$ field, GF is applied.
The lattice gluon propagator, a two-point correlation function, reads
\be
 \langle A^{m}_{\mu}(\widehat{q}\,) A^{n}_{\nu}(\widehat{q}\,') \rangle = \delta^{mn}\Delta_{\mu\nu}(q) L^4 \delta(\widehat{q}+\widehat{q}\,') \ ,
 \label{lattprop}
\ee
where we Fourier transform from the positions $x$ to the momenta $q$.
The transverse and longitudinal SU(3) propagator form factors are \cite{Leinweber:1998im,Oliveira:2012eh},
\bea
 \Delta_\s{\mathrm{T}}(q^2)&=&\frac1{24L^4}\sum_{\mu,\nu,m}(\delta_{\mu\nu}-q_\mu q_\nu/q^2)\langle A^{m}_{\mu}(\widehat{q}\,) A^{m}_{\nu}(-\widehat{q}\,)\rangle,
 \nonumber \\
 \Delta_\s{\mathrm{L}}(q^2)&=&\frac1{8L^4}\sum_{\mu,\nu,m}q_\mu q_\nu/q^2\langle A^{m}_{\mu}(\widehat{q}\,) A^{m}_{\nu}(-\widehat{q}\,)\rangle .
\eea

Analytically, since the $R_\xi$ longitudinal propagator $\Delta_\s{\mathrm{L}}$ remains equal to the tree level one. we should have $q^2\Delta_\s{\mathrm{L}} \equiv\xi$.
The values of $\xi$ chosen are $\xi$ = 0.1, 0.2, 0.3, 0.4, and 0.5. 
Indeed a fit of the data to a constant, show in Fig. \ref{fig:longdress}, yields $\xi=$ 0.103(2), 0.203(2), 0.302(3), 0.402(3) and 0.502(3) respectively,

The $R_\xi$ gluon transverse dressing function $q^2\Delta_\s{\mathrm{T}}$ is fully dynamical and non-perturbative. 
The simulated volume is (3.25 fm)$^4$, large enough to resolve the onset of the non-perturbative effects.  
For comparison, the 
Landau gauge results obtained for a symmetric lattice of $L=80$ and $\beta=6.0$ (gray crosses) are also plotted in Fig. \ref{fig:transdress}.

\begin{figure}[t]
\begin{subfigure}{.45\linewidth}
\includegraphics[scale=0.9,trim={0 0 8.5cm 0},clip]{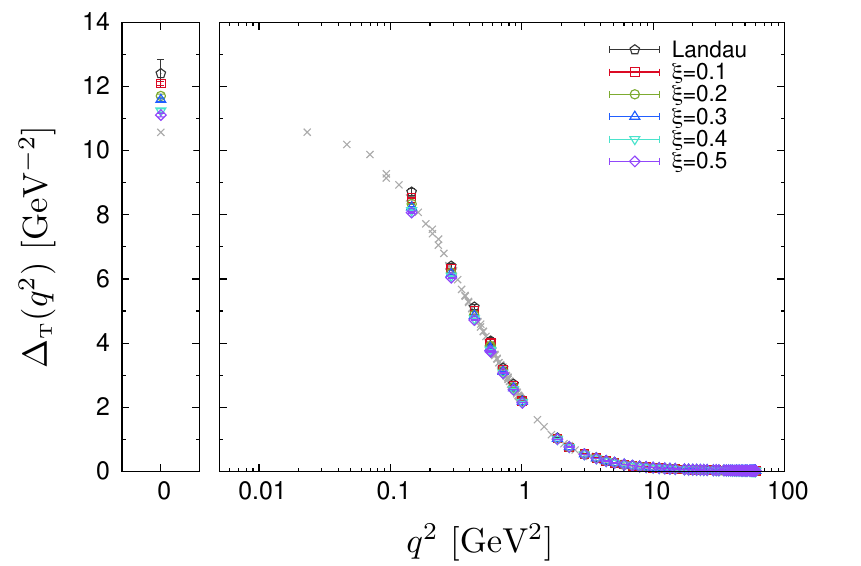}
\caption{ 
\label{fig:transprop}
We show he $R_\xi$ transverse propagator $\Delta_\s{\mathrm{T}}$ (renormalized at $\mu=4.317$ [GeV] ). 
}
\end{subfigure}
\quad
\begin{subfigure}{.45\linewidth}
\includegraphics[scale=0.9,trim={8.5cm 0 0 0},clip]{gl-trans-ratio.pdf}
\caption{
\label{fig:ratio}
Ratio of $R_\xi$ transverse propagators to the Landau gauge transverse propagator.}
\end{subfigure}
\caption{Transverse gluon propagators for different $R_\xi$ gauges. The grey crosses are computed with the Landau gauge for a volume of $80^4$ and provide an estimate for the volume effects expected at $q^2=0$
\cite{Oliveira:2012eh}.}
\end{figure}

We plot the gluon $R_\xi$ transverse propagators in Fig. \ref{fig:transprop}. As in the Landau gauge, we find they show an inflection point, implying that the associated spectral density is not positive definite; it is interpreted as a manifestation of confinement. 
They have a marked tendency to flatten towards the small momentum region, thus providing strong evidence that also in the $\xi\neq0$ case the behaviour of the lower modes of the lattice gluon field are tamed by the dynamical generation of a (momentum-dependent) gluon mass. 
The Landau gauge data on very large volumes suggest that simulations on larger physical volumes suppress the gluon propagator in the infrared region. This would only lead to at most to a decrease of about $\sim 10 \%$ of $\Delta_\s{\mathrm{T}}$ at small momenta for the $R_\xi$ transverse propagator.

We also plot the ratio, in Fig. \ref{fig:ratio}, of the transverse propagator to the Landau gauge propagator $\Delta^{\xi=0}_\s{\mathrm{T}}$ as a function of the momentum for the two values $\xi=0.1$ and $\xi=0.5$.
The data confirms an IR hierarchy such that $\Delta_\s{\mathrm{T}}$ (slightly) decreases for increasing values of the gauge fixing parameter. 
The maximum difference is of $\sim$  10$\%$ for $\xi=0.5$.


\subsection{Continuum studies, gluon mass}

\begin{figure}[t]
\begin{subfigure}{.45\linewidth}
\includegraphics[width=1.07\textwidth]{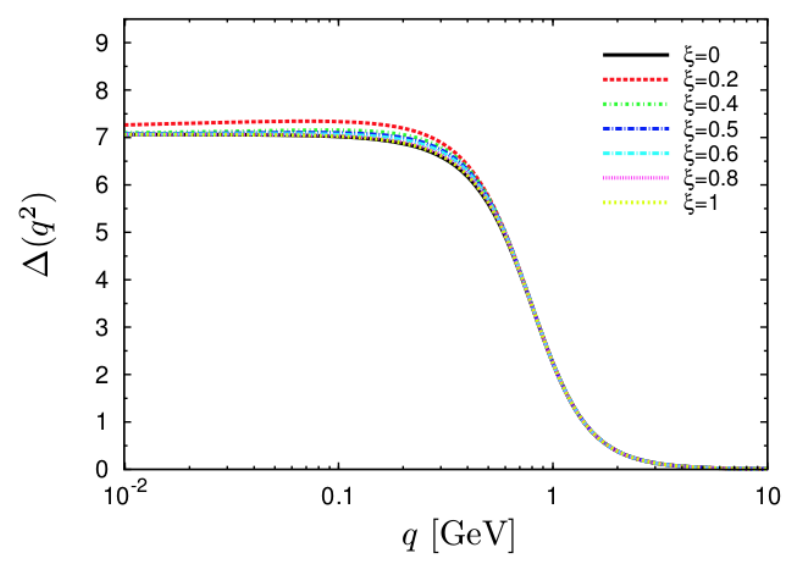}
\caption{
\label{fig:theoretical}
Previous theoretical prediction of the gluon propagator \cite{Aguilar:2015nqa}. }
\end{subfigure}
\qquad
\begin{subfigure}{.45\linewidth}
\vspace{-4pt}
\includegraphics[width=1.03\textwidth]{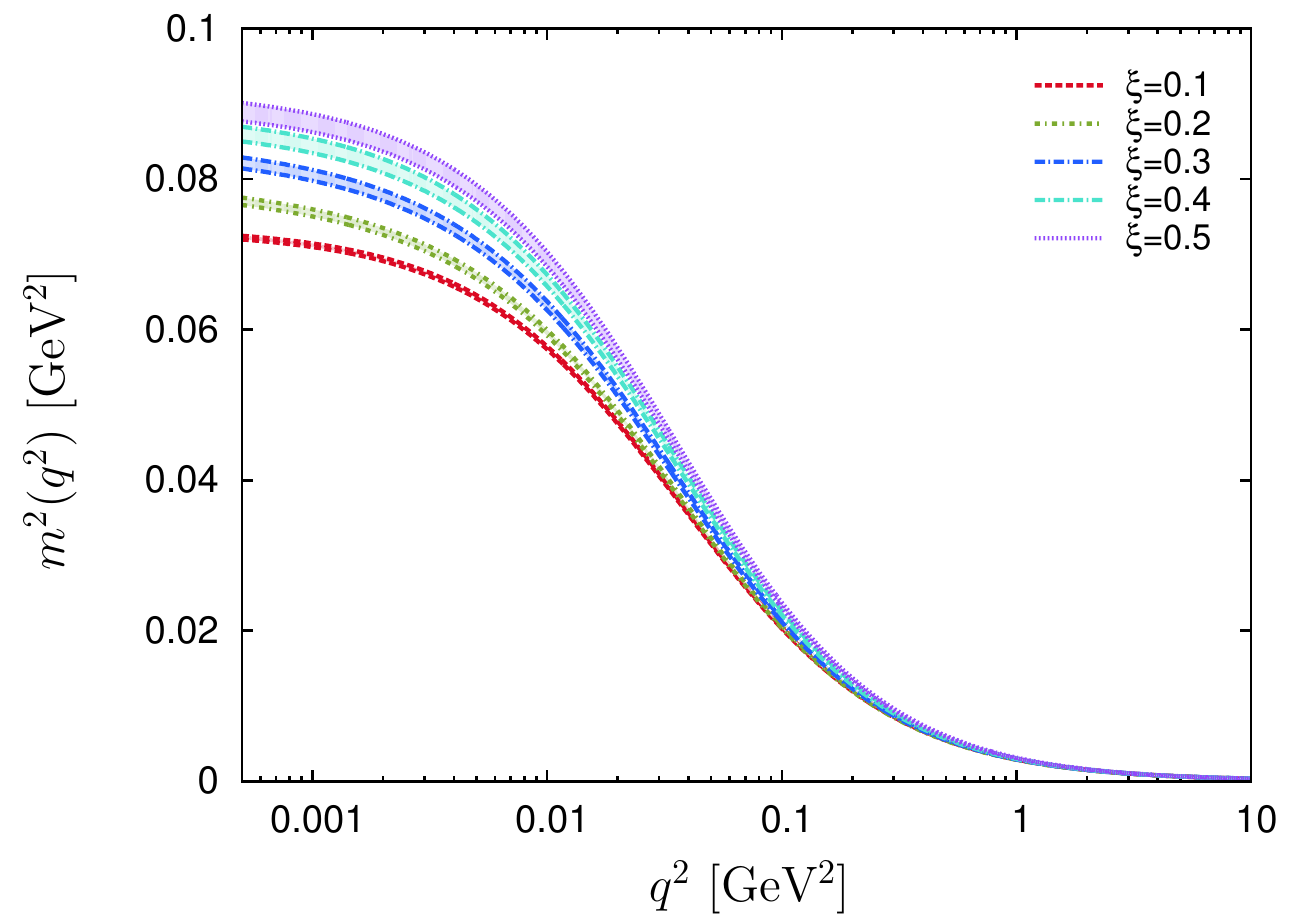}
\caption{
\label{fig:reconstructed}
Reconstructed gluon mass in the $R_\xi$ gauges, using the Nielsen identities of Ref.  \cite{Aguilar:2015nqa}.}
\end{subfigure}
\caption{Comparing with continuum studies and  reconstructing the gluon mass.}
\end{figure}

Our data for the gluon propagator is comparable to the one predicted in Ref. \cite{Aguilar:2015nqa}, shown in Fig. \ref{fig:theoretical},
who include Nielsen identities and a gluon mass.  
Applying the same analysis of of Ref. \cite{Aguilar:2015nqa} to our $R_\xi$ lattice propagators, we find that for small~$\xi$, the $R_\xi$ dynamical mass behaves like 
\be
	m^2(q^2)=\left[a(\xi)+c(\xi)\left(\frac{q^2}{\mu^2}\right)^{\!\!\xi}\log\frac{q^2}{\mu^2}\right]m^2_{\xi=0}(q^2) , 
		\label{resmass}
\ee
where $m^2_{\xi=0}$ is the Landau gauge gluon mass and  to lowest order in the gauge fixing parameter, $a(\xi)=1+a_1\xi$, $c(\xi)=c_\s{\mathrm{NI}}\xi$. 

Using as input our $R_\xi=0$ (Landau gauge), $L=32$, $\beta=6.0$ data for the gluon and ghost propagators we obtain the Landau gauge dynamical mass $m^2_{\xi=0} (q^2)$.
We then solve the renormalization group improved equation  and determine as well as the numerical value $c_\s{\mathrm{NI}}\approx 0.32$. 
We also determine the numerical value of $a_1$ by requiring that the dynamical mass  equals the value of $\Delta_\s{\mathrm T}^{-1}(0)$ for the corresponding value of $\xi$. We obtain $a_1 \approx 0.26$.
Finally including the statistical errors on the lattice propagators, we obtain the gluon running mass for the different $R_\xi$, plotted in Fig. \ref{fig:reconstructed}.


\section{Conclusions and outlook}

From the numerical point of view, our most intensive task
\cite{Bicudo:2015rma,preparation}, is the gauge fixing due to the large number of GF's required and algorithmic issues. 
Up to  large $\xi$'s  and volumes, with a proper combination of various steepest descent methods, we {\em solve the GF in $R_\xi$ gauges} . 
We also compute the lattice SU(3) gluon propagators in $R_\xi$ gauges, for a lattice volume large enough to access the IR dynamics. 
Our $\Delta_\s{\mathrm{T}}(q^2)$ propagators are $\sim$ similar for  $\xi =0.0  -  0.5$ : an inflection point in the few hundreds MeV region and a saturation in the IR. 
 Our propagators are in agreement with very recent continuum analytic studies, and we estimate the dynamically generated $R_\xi$ mass. 
This analysis suggests that dynamical gluon mass generation is a common feature of all $R_\xi$ gauges in SU(3) Yang-Mills theories.

{\bf Acknowledgments}

The authors acknowledge the use of
the GPU servers of PtQCD, supported by CFTP, FCT and NVIDIA; AC33 and Hybrid GPU servers from NCSA's Innovative Systems Laboratory (ISL) and of the computer cluster Navigator, managed by the Laboratory for Advanced Computing, at the University of Coimbra.
P. B. thanks  the ECT{*} Centre and grant FCT UID/FIS/00777/2013.
The work of N. C. is supported by NSF award PHY-1212270. 
P. J. S. acknowledges the support of FCT through the grant SFRH/BPD/40998/2007.


\end{document}